\documentclass[prl, amsmath,superscriptaddress,twocolumn]{revtex4-1}
\usepackage{graphicx}
\usepackage{color}

\begin{document}

\title{Work producing reservoirs: \\
Stochastic thermodynamics with generalized Gibbs ensembles}

\author{Jordan M. Horowitz}
\affiliation{Physics of Living Systems Group, Department of Physics, Massachusetts Institute of Technology, 400 Technology Square, Cambridge, MA 02139}
\author{Massimiliano Esposito}
\affiliation{Complex Systems and Statistical Mechanics, Physics and Materials Science Research Unit, University of Luxembourg, L-1511 Luxembourg, Luxembourg}

\date{\today}

\begin{abstract}  
We develop a consistent stochastic thermodynamics for environments composed of thermodynamic reservoirs in an external conservative force field, that is environments described by the Generalized or Gibbs canonical ensemble.
We demonstrate that small systems weakly coupled to such reservoirs exchange both heat and work by verifying a local detailed balance relation for the induced stochastic dynamics.
Based on this analysis, we help to rationalize the observation that nonthermal reservoirs can increase the efficiency of thermodynamic heat engines.
\end{abstract}


\maketitle 

The noise experienced by small systems is not devoid of form, but has a structure imposed on it by thermodynamics, manifest in the fluctuation-dissipation theorem~\cite{Kubo} and the fluctuation theorems~\cite{Esposito2009,Jarzynski2011}.
This  structure has lead to the formulation of a stochastic thermodynamics that describes the phenomenological relationships among heat, work and entropy fluctuations along individual stochastic trajectories~\cite{Sekimoto,Seifert2012,VandenBroeck2015}.
Stochastic thermodynamics has been wildly successful at systematizing thermodynamic fluctuations in small nonequilibrium systems coupled to one or many thermodynamic reservoirs: macroscopic thermodynamic systems so large they can act as a constant inexhaustible source of energy, particles and/or entropy~\cite{Callen}.
In light of its success, an on-going research endeavor has been to expand the applicability of stochastic thermodynamics to out-of-equilibrium and nonthermal environments as a means to explore the limits of far-from-equilibrium thermodynamics.

While a generic framework for arbitrary environments may be out of reach, there has been success in understanding the exchange of energy and entropy within specific classes of nonequilibrium reservoirs.
For instance, information reservoirs~\cite{Deffner2013,Barato2014c,Horowitz2014b} -- sources of entropy, but not heat -- provide a unified accounting of the thermodynamic costs to operate a Maxwell demon~\cite{Horowitz2014,Parrondo2015}.
In quantum heat engines, quantum nonequilibrium reservoirs~\cite{Leggio2016},  like coherent~\cite{Lutz2009,Scully2003,Quan2006} and squeezed thermal reservoirs \cite{Rossnagel2014,Manzano2016,Huang2012,Correa2014,Long2015}, have been shown to increase the thermodynamic efficiency, sometimes beyond the Carnot limit.
This prediction appears surprising only because the ``hidden'' work necessary to construct the nonequilibrium reservoir has not been accounted for~\cite{Scully2002,Rostovtsev2003}.
Alternative justifications  have been proposed in terms of effective temperatures~\cite{deLiberto2011,Abah2014,Huang2012,Correa2014,Long2015}, generalized thermodynamic forces~\cite{Horowitz2013d} and nonequilibrium entropies~\cite{Manzano2016,Gardas2015}.

To gain perspective on these seemingly remarkable thermodynamic properties of nonequilibrium environments, we develop in this rapid communication a stochastic thermodynamics for a large class of equilibrium environments that display similar behavior: Generalized or Gibbs canonical reservoirs, which are thermodynamic reservoirs in a conservative external force field~\cite{Kardar,Peliti}.
Examples include reservoirs in a fixed electric field or held at constant pressure (instead of volume), as well as moving or rotating reservoirs.
Earlier works on effusions between linearly-translating reservoirs~\cite{Wood2007} and colloidal particles in an external flow~\cite{Speck2008,Lan2015} have demonstrated that such reservoirs modify the thermodynamics.
Our investigation provides a unifying perspective that generalizes these studies. 
We demonstrate that Gibbs reservoirs exchange both heat {\it and} work, much like how a particle reservoir is a source of both heat and chemical work.
This observation challenges the commonly held belief that any energy exchanged with a nonthermal environment is heat~\cite{Scully2003, Abah2014, Lutz2009, Leggio2016,Manzano2016,Huang2012,Correa2014,Long2015}.
In fact, the division between work and heat is intimately connected to the form of the environment.
Based on these observations, we analyze the energetics of a driven spinning paddle in a rotating environment, demonstrating that energy can be extracted from a single reservoir.
We conclude by calculating the maximum efficiency of a cyclic heat engine operating with a Gibbs reservoir, verifying that one can exceed the Carnot efficiency without violating the second law.

To begin, stochastic thermodynamics is a systematic accounting of the random flow of energy and entropy between a small system and its environment.
As such, the fundamental relation that underpins this framework is an equality between the stochastic heat flow ${\dot q}(t)$ into a thermodynamic reservoir and  the entropy flow out of the system ${\dot s}_{\rm e}(t)$~\cite{Seifert2012},
\begin{equation}\label{eq:db}
{\dot s}_{\rm e}(t)=\beta {\dot q}(t),
\end{equation}
where $\beta=1/T$ is the inverse temperature of the reservoir  ($k_{\rm B}=1$).
For thermal or chemical reservoirs, this equality is a consequence of detailed balance or the fluctuation-dissipation theorem~\cite{VandenBroeck2015}.
As such, \eqref{eq:db} is a result of the thermodynamic structure of environmental noise.
To develop a stochastic thermodynamics of Gibbs reservoirs, we will need to demonstrate the validity of \eqref{eq:db} by properly identifying the heat and entropy flow into a Gibbs reservoir.
To this end, we first turn to the macroscopic thermodynamics of the Gibbs ensemble.

The Gibbs canonical ensemble describes a macroscopic equilibrium system with an applied generalized force $F$. 
Its density in phase space $\boldsymbol\zeta=(\boldsymbol\xi,\boldsymbol\nu)$ takes the standard form in terms of the Hamiltonian $H(\boldsymbol\zeta)$~\cite{Kardar,Peliti},
\begin{equation}\label{eq:gibbs}
\rho(\boldsymbol\zeta; F)=e^{-\beta\left(H(\boldsymbol\zeta)-FX(\boldsymbol\zeta)-G\right)},
\end{equation}
where $X(\boldsymbol\zeta)$ is the conserved generalized coordinate conjugate to $F$, and $\beta G(F)=-\ln \int d\boldsymbol\zeta\, e^{-\beta\left(H(\boldsymbol\zeta)-FX(\boldsymbol\zeta)\right)}$ is the (Gibbs) free energy.
A modest list of examples appear in Table~\ref{eq:GibbsTable}.

\begin{table}
\caption{\label{eq:GibbsTable} Gibbs reservoirs with generalized force $F$ and conjugate conserved quantity $X$.  Their product is the work done by the force to prepare the Gibbs state, $W_F=FX$. The four Gibbs reservoirs represented are thermal reservoirs of gas particles of mass $m$ linearly-translating at velocity $V$, rotating at frequency $\boldsymbol\Omega$, held at constant pressure $P$, or constant chemical potential $\mu$.}
\begin{ruledtabular}
\begin{tabular}[t]{lccc}
Ensemble & $F$ & $X$ & $W_F$ \\
\hline
Translating  & ${\bf V}$ & ${\bf p}=m\boldsymbol\nu$ & ${\bf V}\cdot {\bf p}$ \\
Rotating  & $\boldsymbol\Omega$ & ${\bf L}={\boldsymbol\xi}\times {\bf p}$ & $\boldsymbol\Omega \cdot {\bf L}$ \\
$PV$-ensemble & $-P$ & ${\mathcal V}$ & $-P{\mathcal V}$ \\
Chemical &$\mu$ & $N$ & $\mu N$ \\
\end{tabular}
\end{ruledtabular}
\end{table}

In the Gibbs canonical ensemble the internal energy $U$ is not the expectation value of the Hamiltonian $E=\langle H(\boldsymbol\zeta)\rangle$, but is instead~\cite{Landau5}
\begin{equation}
U=\langle H(\boldsymbol\zeta)-FX(\boldsymbol\zeta)\rangle=E-FX.
\end{equation}
One must subtract the work $FX$ done against the external force, which is exactly the energy provided by the external work source to prepare the Gibbs state.
Properly accounting for this work is important, as the entropy is only a function of the internal energy~\cite{Landau5}: 
\begin{equation}
\begin{split}
S(U) &\equiv-\int d\boldsymbol\zeta\, \rho(\boldsymbol\zeta;F)\ln \rho(\boldsymbol\zeta;F) \\
&= \beta(E-FX-G) = \beta(U-G),
\end{split}
\end{equation}
This fundamental distinction has dramatic consequences on the first law of thermodynamics.
Consider an infinitesimal, reversible thermodynamic transformation, where no mechanical work is done apart from the work performed by $F$.
Along this transformation, the Clausius inequality ($Q=TdS$) and the fundamental relation ($dU=TdS$), imply~\cite{Landau5}
\begin{equation}\label{eq:gibbsHeat}
Q=TdS=dU.
\end{equation}
Thus, only the internal energy compensates heat flow.
By contrast, $dE=dU+FdX=Q+W_{F}$ varies due to both heat and the work done by the external force, $W_{F}=FdX$.
For us, this will imply that the energy exchanged between a small system and a Gibbs reservoir during an adiabatic, reversible interaction is not just heat, but must also include work.

We now turn to the stochastic thermodynamics of Gibbs reservoirs.
For clarity of exposition, we focus on
 a paradigmatic example of a small nonequilibrium system: a massive particle of mass $M$ immersed in a dilute gas of particles of mass $m$~\cite{VanKampen}.
If the gas is sufficiently dilute, the stochastic dynamics of the system particle in its phase space $({\bf x},{\bf v})$ is described by the linear kinetic equation for the probability density $P_t({\bf x},{\bf v})$~\cite{VanKampen,VandenBroeck2016}
\begin{equation}
\begin{split}\label{eq:kineticEq}
(&\partial_t+{\bf v}\cdot\nabla_{\bf x}+\frac{1}{M}{\bf f}_t({\bf x})\cdot\nabla_{\bf v})P_t({\bf x},{\bf v})\\
&=\int d\bar{\bf v} \left[W({\bf v}|\bar{\bf v})P_t({\bf x},\bar{\bf v})-W(\bar{\bf v}|{\bf v})P_t({\bf x},{\bf v})\right].
\end{split}
\end{equation}
The left hand side represents a streaming term due to the ballistic motion of the particle under the influence of a time-dependent external force ${\bf f}_t({\bf x})=-\nabla_{\bf x}{\mathcal U}_t({\bf x})+h_t({\bf x})$, with conservative potential ${\mathcal U}$ and nonconservative force $h$.
The ballistic motion is interrupted by collisions with the gas, causing the system's velocity to instantaneously jump ${\bf v}\to\bar{\bf v}$ when the incident gas particle has precisely the right incoming velocity $\boldsymbol\nu({\bf v},\bar{\bf v})$ -- determined from the conservation of kinetic energy and momentum.
As each collision is assumed to be uncorrelated and rare, their probability rate is
\begin{equation}\label{eq:scatter}
W(\bar{\bf v}|{\bf v})=n\sigma(|\boldsymbol\nu-{\bf v}|) \rho(\boldsymbol\zeta ; F),
\end{equation}
where $n$ is the particle density, $\sigma$ is the scattering cross-section, and 
\begin{equation}\label{eq:gibbsRes}
\rho(\boldsymbol\zeta;F) =\frac{1}{Z}e^{-\beta\left(m\boldsymbol\nu^2/2+V(\boldsymbol\xi)-FX(\boldsymbol\zeta)\right)},
\end{equation}
 is the probability to find a gas particle with the appropriate position and velocity, taken to have a Gibbs canonical density with Hamiltonian $H=m{\boldsymbol\nu}^2/2+V(\boldsymbol\xi)$.
The potential $V$ can be left arbitrary, as it does not enter into our analysis.
The conjugate coordinate $X$, however, must be a dynamical variable conserved during the collision for the notion of equilibrium to exist~\cite{Peliti}, that is, we require $X(\bar{\boldsymbol\nu})-X({\boldsymbol \nu})=-(X(\bar{\bf v})-X({\bf v}))$.
An intriguing example is momentum, which we will come back to in our illustrations.

With this setup, we can establish \eqref{eq:db} as a property of Gibbs reservoirs.
Within stochastic thermodynamics, there are two separate methods to identify the entropy flow: the first is the degree of time-reversal symmetry breaking in the dynamics, and the second is from a second-law-like entropy balance~\cite{VandenBroeck2015}.
This discrepancy has lead to an on-going discussion on the proper identification of entropy production~\cite{Spinney2012, Ge2014}.
We will find that there is a symmetry that enforces consistency between the two approaches, just like how the parity symmetry of stationary thermal reservoirs provides the necessary connection~\cite{VandenBroeck2016,Gaspard2013,Deffner2015}.

We first address the approach based on time-reversal symmetry breaking.
To this end, we consider the effect of time-reversal on the dynamics, implemented by reversing the sign of any odd variables, like the velocity $({\bf x},{\bf v})^*=({\bf x},-{\bf v})$.
Under time-reversal the Hamiltonian is symmetric, $H(\boldsymbol\zeta^*)=H(\boldsymbol\zeta)$, which implies that the collisions are as well, $\sigma=\sigma^*$, as they are governed by Hamiltonian dynamics~\cite{Gaspard2013}.
Similarly, the energy of the work reservoir, must also be symmetric, $F^*X(\boldsymbol\zeta^*)=FX(\boldsymbol\zeta)$.

The entropy flow is then determined by the relative likelihood of a stochastic trajectory and its time reverse.
As the ballistic motion between collisions is deterministic, it is symmetric.
We thus focus on the collisions, where the entropy flow per jump ${\bf v}\to \bar{\bf v}$ is given as the ratio of the jump rate $W({\bf v}|\bar{\bf v})$ to its time-reversal $W^*(-\bar{\bf v}|-{\bf v})$ \cite{VandenBroeck2015,VandenBroeck2016},
\begin{equation}\label{eq:flow1}
\Delta s_{\rm e}(\bar{\bf v}|{\bf v})=\ln\frac{W(\bar{\bf v}|{\bf v})}{W^*(-{\bf v}|-\bar{\bf v})}.
\end{equation}
Substituting in \eqref{eq:scatter} followed by \eqref{eq:gibbsRes}, we find that the entropy flow is exactly the change in the Gibbs reservoir's entropy
\begin{align}
\Delta s_{\rm e}(\bar{\bf v}|{\bf v})&=\ln \frac{\rho(\boldsymbol\zeta;F)}{\rho({\bar{\boldsymbol\zeta}}^*;F^*)} \\ \label{eq:ldb2}
&=\frac{\beta}{2}m(\bar{\boldsymbol\nu}^2-\boldsymbol\nu^2)-\beta F(X(\bar{\boldsymbol\nu})-X(\boldsymbol\nu)).
\end{align}
The right hand side represents the stochastic change in the internal energy of the dilute gas during a collision $\Delta u(\bar{\bf v}|{\bf v})$: the kinetic energy change less the work done by $F$.
As this energy is exchanged reversibly, \eqref{eq:gibbsHeat} demands that we equate it to the heat $q(\bar{\bf v}|{\bf v})$:
\begin{equation}\label{eq:ldb}
\Delta s_{\rm e}(\bar{\bf v}|{\bf v}) =\beta\Delta u(\bar{\bf v}|{\bf v})=\beta q(\bar{\bf v}|{\bf v}).
\end{equation}
Thus, when the heat flux is correctly identified with the change in {\em internal} energy, we recover the proper connection between entropy flow and heat.
Alternatively, we have demonstrated a local detailed balance relation for Gibbs reservoirs, as 
\begin{equation}
\ln \frac{W(\bar{\bf v}|{\bf v})}{W^*(-{\bf v}|-\bar{\bf v})}=\beta q(\bar{\bf v}|{\bf v}).
\end{equation}

A second formulation of the entropy flow comes from partitioning the variation of the stochastic Shannon entropy $s(t)=-\ln P_t({\bf x}(t),{\bf v}(t))$ into an irreversible entropy production rate ${\dot s}_{\rm i}(t)$ and an entropy flow $
\Delta s_{\rm e}(\bar{\bf v}|{\bf v})=\ln[W(\bar{\bf v}|{\bf v})/W({\bf v}|\bar{\bf v})]$, distinct from the expression in \eqref{eq:flow1}~\cite{Seifert2012}.
However, there is a symmetry of $W$ following directly from its definition \eqref{eq:scatter} that enforces consistency,
 $W({\bf v}|\bar{\bf v})=W^*(-{\bf v}|-\bar{\bf v})$:
The dynamics induced by the reservoir are symmetric under parity and time-reversal of the external force.

We have now shown that along individual trajectories heat can consistently be identified with entropy flow. 
If we include the change in stochastic Shannon entropy $d_ts(t)=-d_t \ln P_t({\bf x}(t),{\bf v}(t)) $~\cite{Seifert2012}, we arrive at a second law entropy balance \footnote{We use the symbol $d_t$ to denote the total time derivative of state function, which should be contrasted with the over-dot used to signify a flow or current.}
\begin{equation}\label{eq:2Law}
{\dot s}_{\rm i}(t)=d_ts(t)+{\dot s}_{\rm e}(t)=d_t s (t)+\beta {\dot q}(t).
\end{equation}
As a log-ratio of trajectory probabilities, it satisfies a detailed and integral fluctuation theorem and is positive on average~\cite{Esposito2010}, 
\begin{equation}
 {\dot S}_{\rm i}(t)\equiv \langle {\dot s}_{\rm i}(t)\rangle = d_tS(t)+\beta{\dot Q}(t)\ge0,
\end{equation}
where explicitly the heat (or entropy flow),
\begin{equation}\label{eq:avgHeat}
\begin{split}
&\beta{\dot Q}(t)={\dot S}_{\rm e}(t) \\
&\, =\int  W(\bar{\bf v}|{\bf v})P_t({\bf x},{\bf v}) \ln\frac{W(\bar{\bf v}|{\bf v})}{W^*(-{\bf v}|-\bar{\bf v})}\,  d{\bf x}d{\bf v}d\bar{\bf v},
\end{split}
\end{equation}
and entropy change, $S(t)=-\int P_t({\bf x},{\bf v})\ln P_t({\bf x},{\bf v})\, d{\bf x}d{\bf v}$,
\begin{equation}
d_tS(t)=\int W(\bar{\bf v}|{\bf v})P_t({\bf x},{\bf v})\ln \frac{P_t(x,{\bf v})}{P_t(x,\bar{\bf v})}\, d{\bf x}d{\bf v}d\bar{\bf v},
\end{equation}
sum to give the entropy production,
\begin{equation}
{\dot S}_{\rm i}(t) = \int  W(\bar{\bf v}|{\bf v})P_t({\bf x},{\bf v})\ln\frac{W(\bar{\bf v}|{\bf v})P_t({\bf x}, {\bf v})}{W^*(-{\bf v}|-\bar{\bf v})P_t({\bf x}, \bar{\bf v})}\,  d{\bf x}d{\bf v}d\bar{\bf v}
\end{equation}
Take note that the second law only depends on the heat flux into the environment, not the total energy flow.

We have also seen that the energy exchanged with the gas is not just heat, but must also include work.
This directly affects how we account for system energy fluctuations.
Specifically, the energy of our system particle is
\begin{equation}
e(t)=M {\bf v}(t)^2/2+{\mathcal U}_t({\bf x}(t)).
\end{equation}
In between collisions, the motion is deterministic and the energy changes as
\begin{equation}
d_t e(t)=\partial_t {\mathcal U}_t({\bf x}(t))+{\bf v}(t) h_t ={\dot w}(t),
\end{equation}
which is work, as no energy is exchanged with the environment.
In a collision the energy changes discontinuously
\begin{equation}
\Delta e(\bar{\bf v}|{\bf v})=\frac{1}{2}M (\bar{\bf v}^2-{\bf v}^2).
\end{equation}
To relate this to heat and work, we observe that in a collision just the kinetic energy is conserved:
\begin{align}
\Delta e(\bar{\bf v}|{\bf v})&=-\frac{1}{2}m (\bar{\boldsymbol\nu}^2-{\boldsymbol\nu}^2)\\
&=-q(\bar{\bf v}|{\bf v})+F(X(\bar{\bf v})-X({\bf v}))\equiv - {\dot q}(t)+{\dot w}_{F}(t),
\end{align}
where the second line follows from the equality of internal energy and heat \eqref{eq:ldb} (or \eqref{eq:ldb2}) as well as the conservation of $X$.
Thus, in a collision heat {\em and} work are transmitted to the particle.  
Heat comes from the internal kinetic-energy fluctuations of the bath, and work due to the external force.
Combining these observations, we arrive at a first law energy balance for the stochastic energy transfer
\begin{equation}
d_te(t)=-{\dot q}(t)+{\dot w}(t)+{\dot w}_{F}(t).
\end{equation}
On average, we have explicitly
\begin{equation}
d_tE(t)=\langle d_te(t)\rangle = \partial_t \int P_t({\bf x},{\bf v})\left[\frac{1}{2}M{\bf v}^2+ {\mathcal U}_t({\bf x})\right]\, d{\bf x}d{\bf v},
\end{equation}
which is divided as external work
\begin{equation}
{\dot W}=\int P_t({\bf x},{\bf v}) [\partial_t {\mathcal U}_t({\bf x})+{\bf v} h_t ]\, d{\bf x}d{\bf v},
\end{equation}
Gibbs reservoir work 
\begin{equation}
{\dot W}_F=\int W(\bar{\bf v}|{\bf v})P_t({\bf x},{\bf v})[F(X(\bar{\bf v})-X({\bf v}))]\, d{\bf x}d{\bf v}d\bar{\bf v}
\end{equation}
and heat
\begin{equation}
\begin{split}
{\dot Q}=-\int & W(\bar{\bf v}|{\bf v})P_t({\bf x},{\bf v}) \\
&\times\left[\frac{1}{2}M (\bar{\bf v}^2-{\bf v}^2)-F(X(\bar{\bf v})-X({\bf v}))\right] \, d{\bf x}d{\bf v}d\bar{\bf v},
\end{split}
\end{equation}
which is equivalent to \eqref{eq:avgHeat} due to \eqref{eq:ldb}.

A quick example helps to clarify the concepts.
Consider our massive particle confined to one dimension and immersed in a thermal reservoir at inverse temperature $\beta$ moving at a fixed velocity $V$.
In the co-moving frame, Galilean invariance requires the  reservoir to be in equilibrium.
Thus, in that frame a gas particle's velocity $\nu_{\rm cm}=\nu-V$ is distributed according to the Maxwell-Boltzmann distribution~\cite{Landau5}, 
\begin{equation}
\rho(\nu;V)=\sqrt{\frac{\beta m}{2\pi}}e^{-\beta m(\nu-V)^2/2},
\end{equation}
which can be put into the Gibbs form \eqref{eq:gibbs}.
Now according to our analysis the heat exchanged in any  collision is
\begin{equation}
\beta q({\bar v}|v)=\ln \frac{W({\bar v}|v)}{W^*(-v|-{\bar v})}=\frac{\beta}{2}m\Delta(\nu-V)^2,
\end{equation}
which is the change in kinetic energy {\it in the moving frame}.
Thus, heat is only the part of the energy exchanged that goes directly into the internal thermal motion.
The rest is work, $W_F=-Vm\Delta\nu$.
We can re-express the heat and work in terms of system variables, by using the conservation of kinetic energy and momentum: $q=-M\Delta (v-V)^2/2$ and $W_F=VM\Delta v=V\Delta p$.
Heat is determined by looking at the energy exchanged in the moving frame, as was pointed out in \cite{Speck2008} for Brownian particles in an external flow.
Extra energy comes from the ``momentum-work'' due to the exchange of momentum with the translating bath~\cite{Wood2007}.
This work is analogous to the chemical work imparted by a particle reservoir. 

The fact that heat is only due to energy exchanged in the co-moving frame alters what it means to be in equilibrium with a reservoir.
Indeed, if we release our particle in the moving reservoir it will eventually relax to a stationary state flowing with the reservoir, where it exchanges no heat on average;
the particle will be in equilibrium with the moving reservoir.
However, if we trap the particle, say by imposing a harmonic potential ${\mathcal U}(x)=kx^2/2$, the particle will relax to a nonequilibrium steady state characterized by constant dissipation as now the reservoir will be moving relative to the particle.
The dissipation will originate in the work done by the flowing gas as it pushes the particle against the potential gradient.
That work will immediately be dissipated as heat back in the reservoir.
Thus, this steady state is out of equilibrium due to the constant flow of energy from the work source to the heat sink, both in the same Gibbs reservoir.

\begin{figure}
\includegraphics[scale=.4]{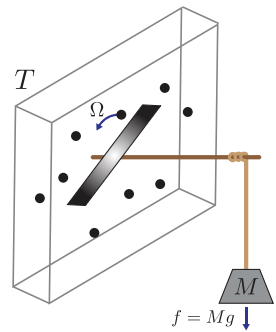}
\caption{Illustration of a paddle in a gas at temperature $T$ rotating with angular velocity $\Omega$.  As the paddle rotates, it lifts a weight of mass $M$ against the gravitational force $f=Mg$.}
\label{fig:feynman}
\end{figure}

We now apply the preceding framework to investigate how to utilize $W_F$ as a resource.
 Our first example, depicted in Fig.~\ref{fig:feynman}, is  
a paddle with moment of inertia $I$ immersed in a dilute gas rotating at frequency $\Omega$.
As the paddle rotates, it raises a mass $M$ against the gravitational force $f=Mg$.
To ease the calculations, we take the diffusive limit of our dynamics, where the impact of each collision is small ($m\ll M$), but their frequency is large ($n\gg 1$).
The resulting dynamics is an underdamped Langevin equation for the angular velocity $\omega$, derived in the Supplemental Material~\cite{suppmat},
\begin{equation}
I\dot{\omega}_t=-f-\gamma(\omega_t-\Omega)+\eta_t,
\end{equation}
where $\gamma$ is the viscosity (obtained from $\sigma$) and $\eta_t$ is zero-mean Gaussian white noise with covariance $\langle \eta_t\eta_s\rangle = (2\gamma/\beta)\delta(t-s)$.
The moving reservoir adds an extra force $\gamma \Omega$, but as this force originates in the environment it alters the definition of heat.

To verify that the rotating bath can indeed lift the weight, we calculate the steady-state energetics.
Details can be found in the Supplemental Material~\cite{suppmat}.
The heat is the energy flux into the bath in the rotating frame, $\omega^{\rm rot}_t=\omega_t-\Omega$: 
\begin{equation}
{\dot Q}=\langle (\omega_t-\Omega)(\gamma(\omega_t-\Omega)-\eta_t)\rangle=f^2/\gamma.
\end{equation}
In addition, the rotation imparts a force $\gamma\Omega$ that does work on the paddle, 
\begin{equation}
{\dot W}_\Omega = -\langle \Omega(\gamma(\omega_t-\Omega)-\eta_t)\rangle=\Omega f.
\end{equation}
The difference is the extracted work 
\begin{equation}
{\dot W}_{\rm ext} = \langle f\omega_t\rangle=f(\Omega-f/\gamma).
\end{equation}
Thus, whenever the rotation is sufficiently strong, $\Omega>f/\gamma$, the work imparted by the reservoir can be usefully extracted.

As a final example, we provide a general analysis of the increase in efficiency for a cyclic heat engine operating between a hot Gibbs reservoir at temperature $T_h$ and cold thermal reservoir at $T_c$.
Over the course of the cycle, $W_{\rm ext}$ work is extracted, $Q_h$ heat enters and $W_F$ work is done on the system, while $Q_c$ heat is exhausted into the cold reservoir.
Efficiency is broadly defined as the ratio of output, $W_{\rm ext}$, to input. 
Here, the energy that enters the system from the hot reservoir comes both as heat $Q_h$ and work $W_F$.
This defines the efficiency as
\begin{equation}
\eta =\frac{W_{\rm ext}}{Q_h+W_F}.
\end{equation}
While this definition of efficiency is formally equivalent to previous studies on engines with nonequilibrium reservoirs, we have refrained from calling the input energy heat.
The ultimate thermodynamic bound on the engine's efficiency is provided by combining the conservation of energy ($W_{\rm ext}=Q_h-Q_c+W_F$), with the second law of thermodynamics, $Q_c/T_c-Q_h/T_h\ge 0$~\eqref{eq:2Law}, which importantly is framed only in terms of heat fluxes:
\begin{align}
\eta= \frac{Q_h+W_F-Q_c}{Q_h+W_F} \le\eta_C+\frac{T_c}{T_h}\left(\frac{W_F}{Q_h+W_F}\right),
\end{align}
where $\eta_C=1-T_c/T_h$ is the Carnot efficiency.
When $W_F/Q_h>0$, we can exceed the Carnot efficiency.  
While the second law restricts the efficiency of heat to work conversion, there is no restriction on work to work conversion.
Indeed, the work from the Gibbs reservoir can be utilized by the engine at 100\% efficiency.

We have argued that Gibbs reservoirs can exchange heat, entropy and work.
This work is an additional thermodynamic resource that can be exploited by thermodynamic engines.
We believe this observation will help rationalize some of the thermodynamic violations observed in devises that utilize nonequilibrium reservoirs.

JMH is supported by the Gordon and Betty Moore Foundation through Grant GBMF4343 and
ME by the National Research Fund Luxembourg through Projects No. FNR/A11/02 and No. INTER/FWO/13/09.

\bibliography{PhysicsTexts.bib,FluctuationTheory.bib,QuantumWork.bib} 

\clearpage

\begin{widetext}

\renewcommand{\theequation}{S\arabic{equation}}
\setcounter{equation}{0}

\section{Supplemental Material}

This supplemental material outlines the derivation of the diffusive limit of the linear kinetic equation [Eq.~(31) of the main text] and lays out its energetics.

\subsection{Derivation of diffusion limit}

We begin with the derivation of the diffusion limit of the linear kinetic equation [Eq.~(4) of the main text] when the mass of the gas particles is small compared to the system particle mass, $m\ll M$, but the collision frequency is large $n\gg 1$.
To make the calculation analytically tractable, we specialize to a one-dimensional system with phase space $(x,v)$ in a thermal reservoir at inverse temperature $\beta=1/T$ linearly translating at velocity $V$.
The linear kinetic equation for $P_t(x,v)$ reads
\begin{equation}
\left(\partial_t+v\partial_x+\frac{1}{M}f_t(x)\partial_v\right)P_t(x,v)=\int d\bar{v} \left[W(v|\bar{v})P_t(x,\bar{v})-W(\bar{v}|v)P_t(x,v)\right],
\end{equation}
with force $f_t(x)=-\partial_x {\mathcal U}_t(x)+h_t(x)$ composed of a conservative potential ${\mathcal U}$ and a nonconservative force $h$.
The transition rates are determined from the conservation of kinetic energy and momentum using scattering theory ~\cite{VanKampen, VanKampen1961, Dorfman},
\begin{equation}
W({\bar v}|v)d{\bar v}=nA\left(\frac{M+m}{2m}\right)^2|{\bar v}-v|\sqrt{\frac{\beta m}{2\pi}}e^{-\frac{\beta m}{2}\left(\frac{M+m}{2m}{\bar v}-\frac{M-m}{2m}v-V\right)^2}d{\bar v},
\end{equation}
where $A$ is the cross-sectional area of the particle.
Here, the gas velocities are sampled from a Maxwell-Boltzmann distribution shifted by the reservoir velocity $V$.
Now, as the mass of the gas particles becomes small, the changes in velocity of the particle in a collision may change appreciable; however, the momentum changes will become small. 
This suggest that a diffusion limit emerges only for the momentum $p=Mv$~\cite{VanKampen1961}:
\begin{equation}
W({\bar p}|p) d{\bar p} =\frac{nA}{M^2} \left(\frac{M+m}{2m}\right)^2 |{\bar p}-p|\sqrt{\frac{\beta m}{2\pi}} e^{-\frac{\beta m}{2}\left(\frac{M+m}{2mM}{\bar p}-\frac{M-m}{2mM}p-V\right)^2}d{\bar p}
\end{equation}

We now show that these jump dynamics are well approximated as a continuous diffusion process in the limit where the mass of the gas particles is small $m\ll M$ and their density is large $n\gg 1$, so that the frequency of collisions is very high, yet each collision only causes a small change in the momentum.
To this end, we introduce a small parameter through the scaling
\begin{equation}
\epsilon=m/M\ll1,\qquad n=\rho/\epsilon^{1/2} \gg 1,
\end{equation}
for an effective, scaled density $\rho$.
We will see that these particular scaling relationships are consistent with a diffusive limit.
With these definitions the transition rates become
\begin{align}
W({\bar p}|p)d{\bar p} &= \frac{A}{M^2}\frac{\rho}{\epsilon^{1/2}} \left(\frac{1+\epsilon}{2\epsilon}\right)^2 |{\bar p}-p|\sqrt{\frac{\beta M\epsilon}{2\pi}} e^{-\frac{\beta}{2 M\epsilon}\left(\frac{1+\epsilon}{2}{\bar p}-\frac{1-\epsilon}{2}p-\epsilon M V\right)^2}d{\bar p} \\ \label{eq:Weps}
&\approx \frac{\rho A}{M}\frac{1}{(2\epsilon)^{3/2}} |{\bar p}-p|\sqrt{\frac{\beta}{8\pi M\epsilon}} e^{-\frac{\beta}{8 M\epsilon}\left({\bar p}-p+2\epsilon(p-MV)\right)^2}d{\bar p},
\end{align}
where in the second line we have made a slight rearrangement, keeping only the dominant nontrivial behavior in $\epsilon$.
We see here that roughly the typical changes in momentum are highly peaked around $\Delta p={\bar p}-p = 2\epsilon(p-V)$, with a width $\sim\epsilon$.
Thus, each collision only changes the momentum a little, as desired.
The rate of collisions $\sim \epsilon^{-3/2}$, so that their likelihood grows with decreasing $\epsilon$.
Following Gardiner, transition rates of the form in \eqref{eq:Weps}, lead to a Fokker-Planck equation for the diffusive dynamics, if the following three quantities behave accordingly~\cite{Gardiner}:
\begin{align}\label{eq:a0}
\alpha_0(p) &= \int d{\bar p}\, W({\bar p}|p) = \Upsilon/\epsilon \\ \label{eq:a1}
\alpha_1(p) &= \int d{\bar p}\, ({\bar p}-p)W({\bar p}|p) = A(p)\Upsilon \\ \label{eq:a2}
\alpha_2(p) &= \int d{\bar p}\, ({\bar p}-p)^2W({\bar p}|p),
\end{align}
where $\Upsilon$ and $A(p)$ are defined by these equations.
Then $\alpha_1$ and $\alpha_2$ are the drift and diffusion coefficients of the Fokker-Planck equation, respectively, \emph{i.e.},
\begin{align}
\int d\bar{p} \left[W(p|\bar{p})P_t(x,\bar{p})-W(\bar{p}|p)P_t(x,p)\right] \approx  -\partial_p\alpha_1(p)P_t(x,p)+\frac{1}{2}\partial_p^2\alpha_2(p)P_t(x,p).
\end{align}
In the following, we calculate $\alpha_0,\alpha_1,\, {\rm and}\, \alpha_2$.

We start with $\alpha_0$:
\begin{align}
a_0(p) &= \frac{\rho A}{M}\frac{1}{(2\epsilon)^{3/2}} \sqrt{\frac{\beta}{8\pi M\epsilon}} \int |{\bar p}-p|e^{-\frac{\beta}{8 M\epsilon}\left({\bar p}-p+2\epsilon(p-MV)\right)^2}d{\bar p},
\\ \label{eq:a0int}
&\approx \frac{\rho A}{M}\frac{1}{(2\epsilon)^{3/2}} \sqrt{\frac{\beta}{8\pi M\epsilon}} \int |{\bar p}-p|e^{-\frac{\beta}{8 M\epsilon}\left({\bar p}-p\right)^2}d{\bar p},
\end{align}
where in the second line we have dropped the $\epsilon$-dependence on the mean of the Gaussian, which leads to a higher-order correction in $\epsilon$.
The resulting integral is analytically tractable, with the result
\begin{equation}
\alpha_0(p) = \frac{1}{\epsilon}\frac{\rho A}{(\beta M\pi)^{1/2}} \equiv \Upsilon/\epsilon,
\end{equation}
which has the appropriate scaling in \eqref{eq:a0}.

Next we determine $\alpha_1$:
\begin{align}
\alpha_1(p)&=\frac{\rho A}{M}\frac{1}{(2\epsilon)^{3/2}} \sqrt{\frac{\beta}{8\pi M\epsilon}} \int ({\bar p}-p) |{\bar p}-p|e^{-\frac{\beta}{8 M\epsilon}\left({\bar p}-p+2\epsilon(p-MV)\right)^2}d{\bar p} \\
&\approx\frac{\rho A}{M}\frac{1}{(2\epsilon)^{3/2}} \sqrt{\frac{\beta}{8\pi M\epsilon}} \int ({\bar p}-p) |{\bar p}-p|e^{-\frac{\beta}{8 M\epsilon}\left({\bar p}-p\right)^2}\left(1+\frac{\beta}{8M}2(p-MV)({\bar p}-p)\right)d{\bar p}
\end{align}
where in the second line we expanded the exponent to lowest order in $\epsilon$.
The integral can be performed analytically, with the result
\begin{equation}
\alpha_1(p) = -2(p-V)\Upsilon \equiv A(p)\Upsilon.
\end{equation}

Lastly, we calculate $\alpha_2$:
\begin{align}
\alpha_2(p)&=\frac{\rho A}{M}\frac{1}{(2\epsilon)^{3/2}} \sqrt{\frac{\beta}{8\pi M\epsilon}} \int ({\bar p}-p)^2 |{\bar p}-p|e^{-\frac{\beta}{8 M\epsilon}\left({\bar p}-p+2\epsilon(p-MV)\right)^2}d{\bar p}\\
&\approx \frac{\rho A}{M}\frac{1}{(2\epsilon)^{3/2}} \sqrt{\frac{\beta}{8\pi M\epsilon}} \int ({\bar p}-p)^2 |{\bar p}-p|e^{-\frac{\beta}{8 M\epsilon}\left({\bar p}-p\right)^2}d{\bar p} \\
&=\frac{2M}{\beta}\Upsilon
\end{align}

Putting these results to gather, we have for the evolution of the probability density $P_t(x,p)$
\begin{align}
\left(\partial_t+\frac{p}{M}\partial_x+f_t(x)\partial_p\right)P_t(x,p)=\partial_p\left[2(p-MV)\Upsilon P_t(x,p)\right]+\frac{1}{2}\partial_p^2\left(\frac{2M}{\beta}\Upsilon\right)P_t(x,p).
\end{align}
To put this is a more recognizable form, we switch to the velocity $v=p/M$ and formally introduce a viscosity $\gamma=2\Upsilon/M$,
\begin{equation}
\partial_tP_t(x,v)=-v\partial_xP_t(x,v)-\frac{f_t(x)}{M}\partial_vP_t(x,v) + \frac{\gamma}{M}\partial_v (v-V)P_t(x,v)+\frac{\gamma}{\beta M^2}\partial_v^2P_t(x,v).
\end{equation}
This Fokker-Planck equation represents the evolution of an underdamped Brownian particle in a thermal reservoir at inverse temperature $\beta$ in an external force $f+\gamma V$.
Notice, the force $\gamma V$ originates in the bath, and therefore must be interpreted as part of the viscous damping.
For completeness, the equivalent Langevin equation is~\cite{Gardiner}
\begin{align}
{\dot x}_t&=v_t \\ \label{eq:LangevinV}
M{\dot v}_t&=f_t(x_t)-\gamma(v_t-V)+\xi_t
\end{align}
where $\xi_t$ is Gaussian white noise of zero mean with covariance $\langle \xi_t\xi_s\rangle=(2\gamma/\beta)\delta(t-s)$ satisfying the fluctuation-dissipation theorem.
The equivalent equation for rotations appears as Eq.~(31) of the main text.

\subsection{Energetics}

We now turn to the average energetics in the diffusive limit. 
The division of the variation of energy into work and heat is determined by the definition of heat, so we address heat first.
The average heat exchanged is
\begin{equation}\label{eq:heat}
{\dot Q} = \int d{\bar v}\,  J({\bar v}|v) \Delta q({\bar v}|v),
\end{equation}
where the velocity current
\begin{align}
J({\bar v}|v) = W(v|\bar{v})P_t(x,\bar{v})-W(\bar{v}|v)P_t(x,v)
\end{align}
arises solely from the jumps in velocity caused by gas collisions, and 
\begin{equation}
\Delta q({\bar v}|v)= -\frac{1}{2}M\left(({\bar v}-V)^2-(v-V)^2\right)
\end{equation}
is the heat exhausted into the reservoir per jump [Eq.~(30) of the main text].

Now the diffusive limit allows us to approximate the average heat flux as follows.
Our analysis of the limiting jump dynamics gives the approximate velocity current as
\begin{align}\label{eq:currentApprox}
J({\bar v}|v) \approx - \frac{\gamma}{M}\partial_v (v-V)P_t(x,v)+\frac{\gamma}{\beta M^2}\partial_v^2P_t(x,v)\equiv J(v),
\end{align}
which is the average velocity in the frame sliding along with the reservoir at velocity $V$.
The heat per jump is approximately
\begin{align}\label{eq:heatApprox}
\Delta q({\bar v}|v)\approx - M(v-V).
\end{align}
Thus, substituting \eqref{eq:currentApprox} and \eqref{eq:heatApprox} into \eqref{eq:heat}, gives the average heat flux in the diffusive limit as
\begin{equation}\label{eq:avgHeatApprox}
{\dot Q} = \int dv\, J(v) M(v-V)=\langle (v_t-V) \left[\gamma(v_t-V)-\xi_t\right]\rangle,
\end{equation}
where the average on the right hand side is the equivalent result in the Langevin picture.
The product must be interpreted in the Stratonovich sense~\cite{Sekimoto}.
We see the heat is produced by the dissipative forces in the moving frame.

With the identification of the stochastic heat flux (cf.~\eqref{eq:avgHeatApprox})
\begin{equation}\label{eq:LangevinHeat}
{\dot q}(t) = (v_t-V) \left[\gamma(v_t-V)-\xi_t\right],
\end{equation}
we can address the stochastic energy balance by analyzing the time derivative of the energy $e(t)=Mv_t^2/2+{\mathcal U}_t(x_t)$:
\begin{align}
d_t e(t) &= Mv_t{\dot v}_t+\partial_t{\mathcal U}_t(x_t)+v_t\partial_x{\mathcal U}_t(x_t) \\
&=v_t\left(-\partial_x{\mathcal U}_t(x_t)+h_t-\gamma(v_t-V)+\xi_t\right)+\partial_t{\mathcal U}_t(x_t)+v_t\partial_x{\mathcal U}_t(x_t) \\
&= -v_t(\gamma(v_t-V)-\xi_t)+\partial_t{\mathcal U}_t(x_t)+v_th_t,
\end{align}
where we have used the Stratonovich interpretation and in the second line have substituted in the Langevin equation for $v_t$ \eqref{eq:LangevinV}.
To make contact with the heat \eqref{eq:LangevinHeat}, we arrange as
\begin{align}
{\dot e}(t)= \underbrace{-(v_t-V)(\gamma(v_t-V)-\xi_t)}_{-{\dot q}(t)}+\underbrace{V(\gamma(v_t-V)-\xi_t)}_{{\dot w}_V}+\underbrace{\partial_t{\mathcal U}_t(x_t)+v_th_t}_{\dot w},
\end{align}
after identifying the heat flux into the environment ${\dot q}$, the work done by the motion of the reservoir ${\dot w}_V$ and the work done by the forces on the particle ${\dot w}$.
The equivalent definitions for a rotating paddle are used to calculate its energetics following Eqs.~(32)-(34) of the main text.
\clearpage
\end{widetext}

\end{document}